# To PLAnetary Transit Or NOT?
# An extremely large field of view camera with a CaF$_2$ component tested in thermo-vacuum

M. Bergomi*[a,b], D. Magrin [a], J. Farinato [a], V. Viotto [a], R. Ragazzoni [a], A. Brunelli [a], M. Dima [a], P. Christiansen [c], M. Ghigo [d], D. Laubier [e], H. Pasquier [e], D. Piazza [c], I. Pagano [g], G. Piotto [b], G. Basile [h], C. Catala [i]

[a]INAF-Osservatorio Astronomico di Padova, Padova, Italy
[b] Dip. Fisica e Astronomia, Università degli Studi di Padova, Italy
[c] Physikalisches Institut, University of Bern, Switzerland
[d]INAF-Osservatorio Astronomico di Brera, Merate, Italy
[e]CNES, Toulouse, France
[g]INAF-Osservatorio Astrofisico di Catania, Italy
[h]SELEX Galileo S.p.A., Firenze, Italy
[i]Observatoire de Paris-Meudon, France

**ABSTRACT**

Because of its nicely chromatic behavior, Calcium Fluoride (CaF$_2$) is a nice choice for an optical designer as it can easily solve a number of issues, giving the right extra degree of freedom in the optical design tuning. However, switching from tablet screens to real life, the scarcity of information -and sometimes the bad reputation in term of fragility- about this material makes an overall test much more than a "display determination" experiment. We describe the extensive tests performed in ambient temperature and in thermo-vacuum of a prototype, consistent with flight CTEs, of a 200mm class camera envisaged for the PLATO (PLAnetary Transit and Oscillations of Stars) mission. We show how the CaF$_2$ lens uneventfully succeeded to all the tests and handling procedures, and discuss the main results of the very intensive test campaign of the PLATO Telescope Optical Unit prototype.

## 1. INTRODUCTION

Optical instrumentation for space applications has to deal with constraints coming from the space environment, such as thermal, mechanical (vibration, acceleration, shock) and radiation loads, vacuum conditions, other than on maximum exploitation of available payload mass. These factors combined with long lifetimes in space and often demanding optical performance requirements, are driving elements in the design of space optics and in the selection of appropriate materials and technology [1].

Calcium fluoride (CaF$_2$) is extremely appreciated by optical designers for the advantages it offers compared to more commonly used materials. It is a colorless crystal with a fluorite-type crystalline structure, which provides excellent transmission along with very low dispersion over a wide wavelength range (from UV to mid IR 0.130 to 9 um, according to [2]). This means that a single lens can be used to keep chromatism under control, instead of a combination of lenses (doublets, triplets…) with a resulting mass reduction, a prime importance factor in space systems. Moreover, it is naturally resistant to high radiation [3][4], becoming an even more desirable material in a highly irradiated mean as space.

However, CaF$_2$ has rarely been considered for space systems due to issues concerning the fragility of crystal-lenses and its high susceptibility to temperature changes, because of its very high CTE, which struggle against the choice of this material for space missions, where temperature excursion and high vibration occur, in particular, during launch phase.

---

* maria.bergomi@inaf.oapd.it; phone: + 39 049 8293428

Moreover, due to its brittleness, until a few years ago it was extremely difficult machining it achieving a high-quality surface, which is directly correlated to its optical performance [5].

Considering, though, the great advantages provided by the use of this material, along with excellent processing possibilities for being one of the hardest materials in the fluoride crystal family and great improvements on its machining thanks to extensive studies conducted in recent years due to its use as prime material in semi-conductor lithography, and in the light of encouraging results coming from recent tests performed on $CaF_2$ blanks exposed to space environment conditions (high variation of temperature, UV and Gamma radiation) [6], it has been chosen in final PLATO (PLAnetary Transit and Oscillations of Stars) Telescope Optical Unit (TOU) design to be used as material for the optical system stop and a specific mechanical structure has been designed to hold the lenses minimizing effects due to launch stress.

In this paper will be briefly presented the opto-mechanical design of the Telescope Optical Unit prototype and thoroughly described its internal alignment performed at INAF-Padova laboratories and main results of thermo-vacuum test of the aligned optical system, carried out at SELEX Galileo SpA facilities. Behavior of CaF2 during TOU performance verification will be highlighted, as well as results of specific tests performed on CaF2 blanks with the aim to show the response of this material when subjected to thermal and shock stress comparable to a launch through a Soyuz-Fregat type vehicle.

## 2. PLATO

PLATO [7] was selected in 2010 as one of the 3 medium-class mission for definition study in the framework of the ESA Cosmic Vision 2015-2025 program.

It was proposed as the next generation planetary transit experiment with the aim to detect and characterize exoplanetary systems in the solar neighborhood, including both transiting planets down to low-mass earth-size planets in the habitable zone, and their host stars. Focusing its observations on a large sample of bright stars ($m_V$< 14), it would have allowed to measure radii, masses and ages of this planets with high accuracies (≈1%) [7].

Its concept is based on a multi-instrument approach, composed of 34 small Telescope Optical Units (TOUs) with a very wide Field of View (FoV), grouped in a way to allow the best "overlapping line of sight" [8][9], reaching an extreme wide-field capability (total FoV 2180 $deg^2$) over a wide wavelength range. Two of the TOUs, were designed to be fast telescopes, observing observe vey bright stars (mV< 8) trough broad-band filters.

All main parameters, which drove the TOU optical and mechanical design, leading to a the 6-lenses all refractive system, with 4 CCDs mounted in mosaic in the Focal Plane Array (FPA), are listed in Table 1.

Unluckily PLATO was not selected as one of the two projects chosen for the 2017 and 2019 launch opportunities (M1, M2), but, nevertheless, we want to share with the community the lessons learned from the PLATO TOU prototype assembly integration and test phase, focusing on $CaF_2$ performance.

| Spectral range | 500 – 1000 nm |
|---|---|
| Entrance Pupil Diameter | 120 mm |
| Working F/# | 2.06 @ 700 nm |
| Field of View (1) | 151.5 $degree^2$ |
| Image quality | 90% Enclosed energy < 2 x 2 $pixel^2$ |
| Plate scale | 15 arcsec/pixel |
| Pixel size | 18 micron |
| CCD format | 4510 x 4510 (x 4) $pixel^2$ |
| Optical elements weight | 5 Kg |
| Working Temperature | -80°C |
| Working Pressure | 0 atm |

Table 1: General performance and main parameters of the baseline optical configuration [2], [4]

# 3. TELESCOPE OPTICAL UNIT PROTOTYPE

A TOU prototype BreadBoard (BB, hereafter) has been designed and produced during the phase A of the project, to validate assembly integration and test procedures, verify their time-compatibility with respect to industry framework (since the realization of the 34 TOUs was intended to be assigned to industries), and test the on-axis system performance in ambient (20 °C, 1 atm) and thermo-vacuum conditions (-80°C, 0 atm).

BB consists of a set of 6 custom made lenses as close as possible to final design and a mechanical structure equivalent to TOU in terms of thermal behavior.

## 3.1 Optical Design

Since the aim of the BB was to check the on-axis optical quality, some elements have been modified with respect to final TOU design [10], to facilitate and speed up the manufacturing process. In the BB design, for instance, the entrance window was not inserted and lenses present differences in surfaces curvatures, thicknesses and aperture with respect to the baseline optical design. The first two lenses (L1 and L2) have been modified in order to avoid the aspheric terms on the first surface of L1 and to maintain a suitable optical quality only in the center field. These two lenses have been named L1* and L2*. For procurement reasons the glass of L5, KZFSN5, has been substituted by N-KZFS5. All breadboard lenses are standard spherical surfaces, manufactured by SESO, and the first surface of the first of the third lens, the CaF2 lens, is close to the optical system stop and guarantees a real entrance pupil diameter of 120 mm. A layout design of the breadboard is shown in Figure 1 and lens parameters are reported in Table 2.

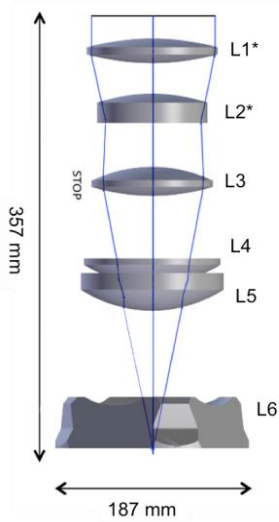

| | | R (mm) | Thickness (mm) | Glass | Shape | Diameter (mm) |
|---|---|---|---|---|---|---|
| L1* | S1 | 269.675 | 21.000 | S-FPL51 | circular | 126.2 |
| | S2 | -269.675 | 26.206 | | | 125.0 |
| L2* | S3 | 187.390 | 7.000 | N-KZFS11 | circular | 106.4 |
| | S4 | 76.000 | 55.520 | | | 98.6 |
| L3 | S5 | 146.735 | 25.000 | CAF2 | circular | 114.8 |
| | S6 | -224.930 | 56.751 | | | 117 |
| L4 | S7 | -686.930 | 19.000 | S-FPL53 | circular | 130.4 |
| | S8 | -137.440 | 14.548 | | | 131.6 |
| L5 | S9 | -100.025 | 8.000 | N-KZFS5 | circular | 130.6 |
| | S10 | -140.760 | 109.954 | | | 137.6 |
| L6 | S11 | -100.025 | 8.000 | N-BK7 | square | 144.0 |
| | S12 | infinity | 5.948 | | | 186.6 |

Figure 1: Layout of breadboard optical design     Table 2: Main lens parameters of breadboard design

The expected optical performances indicated through Enclosed Energy (EE) at ambient and working temperatures with HeNe source are shown in Figure 2. In the case of the ambient temperature, the 90% EE is largely inside 4 TOU pixels (TOU pixel-size: 18 micron) while at the working temperature it is by far inside 2 TOU pixels and it is maintained for a field of view of ±1.38 arcmin. The optical quality of the system had to be checked by using a CCD camera having a pixel size of 7.4 micron, allowing an oversampling of the PSF. The best focal plane location with respect to the rear surface of L6, in the ambient temperature case and in the working temperature case, is 5.948 mm and 4.000 mm respectively.

Simulations have been performed to verify the effects of the radiation environment surrounding the system (PLATO was planned to be located at Lagrangian point L2). These radiations contribute to the degradation of the materials and components of the satellite during the mission. In particular this study has been focused on the radiation damage mechanisms in glasses so to investigate if the present proposed materials for the lenses of the TOUs are usable for the manufacturing of the optical components. Results over $CaF_2$ lens have confirmed its properties as a natural radiation hardened glass [3][4]: no reduction in transmittance was observed after simulated exposure to protons having 5 MeV energy. Considering an irradiation of about 27000 rad over PLATO lifetime (8 years), it is expected that this lens should maintain the transmittance to the level of 96% or better over the whole range, while the other lenses subjected to similar irradiations, showed a decrease in transmittance of about 20-25%, varying for each wavelength [10].

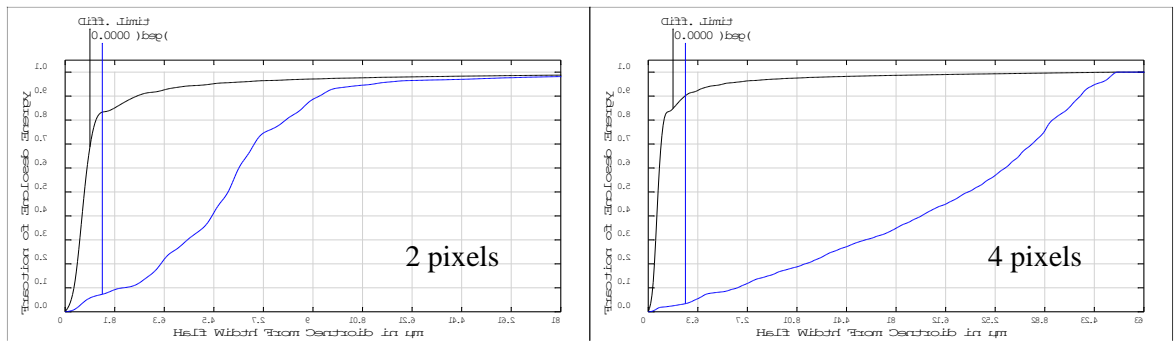

Figure 2: Expected Enclosed Energy for the on axis field of the breadboard in the case of working temperature (left) and ambient temperature (right).

## 3.2 Mechanical Design

The use of materials with different coefficient of thermal expansion and the large temperature difference between integration and operation requires a design able to accommodate the dimensional changes due to CTE of the assembled components without leading to mechanical stresses that could result in structure warping (loss of positioning accuracy) or even in structural failures, especially in the case of brittle optical material, such as $CaF_2$. In addition there are stringent requirements with respect to the positioning of the optical elements. In facts the allowed relative displacements are in the range of 10-100 μm.

The BB mechanical structure is a simplified structure, equivalent to the final AlBeMet one, in terms of thermal expansion coefficient. The chosen Iron & Nickel alloys have a nearly identical CTE (TOOLOX 44 (Plates): 13.5; Inconel 601 (Columns): 13.75 vs AlBeMet :13.9).

The machining tolerances of the mechanical components and of the lens-mount subassemblies can be corrected during integration by shimming. Therefore all elements directly bolted to the TOU structure have a three-point attachment, whereby spherical washers are foreseen to minimize mechanical stresses when the screws are tightened. Flexible, quasi-isostatic mounts were therefore implemented at each mechanical interface, in particular for BB, lenses to lens mounts and lens mounts to TOU structure.

For $CaF_2$ lens mount has been chosen a Dispal S232 material, a powder metallurgical aluminum alloys showing very good CTE match to the glass and presenting a good thermal conductivity.

The lenses were glued to their flexible quasi-isostatic mounts (shown in Figure 3), with the bi-component glue 3M Scotch-Weld 2216 and an adhesive activator, operation performed at SELEX Galileo premises. Each mount is equipped with small shelves, on which the lens lies during the gluing operations in proper position, except for L6, whose pads are positioned in the middle of the lens four lateral faces.

The gluing of the lenses has been performed in order to avoid the glue to act as an interface between the mount and the lens itself and in a way to ensure a centering precision of the order of 1/10 of a millimeter, value which is well below the travel adjustments of each mount, which is of the order of 1mm, thus allowing the proper centering during the alignment of the lenses in the BB (see Section 4). Tilt is also not an issue, since shimming is anyway required during the lenses alignment for the tilt compensation. The flatness of the mechanical reference of the mounts is of the order of 50μm, corresponding to a maximum tilt, at the level of the smallest lens (~114mm of diameter), of about 90"; this number is comparable to the measured wedges of the lenses of the BB lenses and the latter has anyway to be compensated by shimming, being the preliminary tilt alignment tolerances of the lenses of the order of 30".

The breadboard mechanical structure differs from the final model one for thermal conductivity, specific heat and density. For these reasons, while tests at a given temperature are significant, the breadboard is not suitable for testing when temperature gradient occurs. Therefore the aligned breadboard will be tested at ambient temperature and at the final working temperature of -80°C.

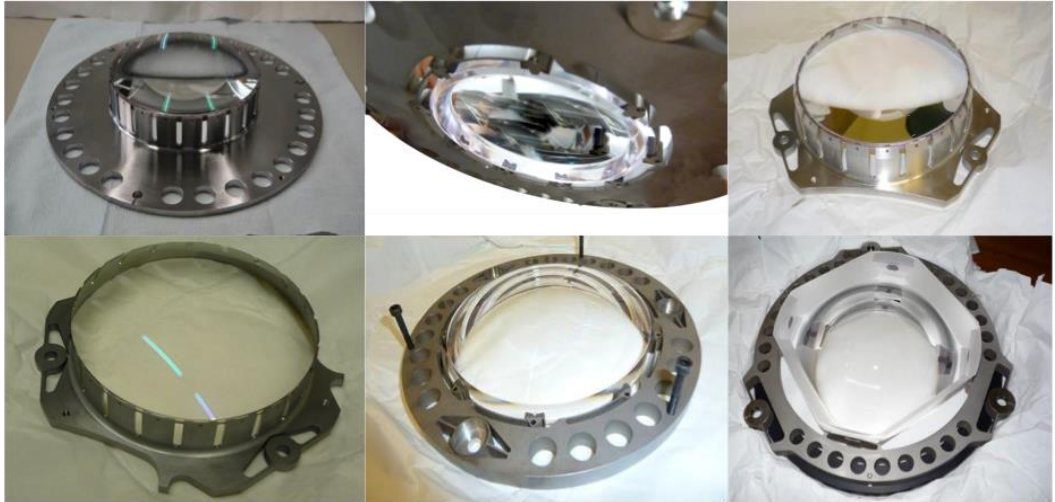

Figure 3: BB lenses glued on their mounts. From top left to bottom right: L1, L2, L3, L4, L5 and L6

## 4. TOU PROTOTYPE ALIGNMENT

To allow an easier assembly of the lenses inside the structure and the centering of the mounts during the alignment of the system, the TOU mechanical structure was kept in vertical, to take advantage from gravity. In these phases a handling structure was required, to hold the TOU mechanical structure vertically and allow a 180° rotation of the entire telescope, in order to insert the lens mounts always from the top (Figure 7).

The opto-mechanical setup concept devised to align the breadboard is shown in Figure 5.A laser beam is expanded using an objective, a spatial filter and a collimating lens. A flat mirror folds the light upward, and a variable iris is used to select the proper diameter, which depends on spots visibility. A beam-splitter sends the light toward a flat mirror that sends the beam toward BB optical axis and allows the light back-reflected from the BB lenses to be observed on a CCD test camera (CCD #1, hereafter). The lenses were inserted one at a time in the optical path and aligned to the reference laser, looking at back-reflected spots and Newton rings, when visible, on CCD#1. A repositionable corner-cube and variable iris could be inserted in the optical path between the beam-splitter and the flat mirror to define a reference position. In order to maximize the contrast and the Airy and Newton rings visibility, thin light shields were inserted between lenses, when needed, to isolate the back reflected light coming only from that lens. Another test CCD (CCD#2, hereafter), located after the BB, collects the transmitted beam, used as reference when a new lens is inserted.

The fixing holes on the mounts were large enough to allow a centering travel of about ±1mm, and the position of the lens was fixed only after the optical alignment tightening the screws connecting the mount to the main structure. To maximize the mechanical precision during the lens centering and to avoid any mount shift during the fixing operation, a dedicated tool was designed. Four bars are laterally fixed to the main mechanical structure; two of them, fixed at a 90 degrees separation, are provided with a micrometric screw in correspondence of each lens mount, to adjust the lens position, while the other two bars have pre-loaded systems acting as a piston to keep the lens mount always in contact with the micrometric screws (see Figure 4).

The first step for the BB alignment was the reference definition on the CCD #1 (pixel-size=6.45 μm) which would have then recorded the back-reflected spots, coming from the BB lenses. Due to the indetermination of the positioning of a repositionable corner-cube and iris, inserted in the beam, in addition to the precision in defining the position of the spot back-reflected from the corner-cube, an overall indetermination of ± 8 pixels is obtained.

The opto-mechanical design of the TOU introduced some constraints concerning the order in which the lenses had to be inserted in the TOU structure. Lens L3, the closest to the system physical pupil, first lens to be inserted in the optical path and since it was not equipped with any centering or tilting adjustment facility, it was mounted on the mechanical structure centered with mechanical precision. The laser beam, used as a reference for the lenses alignment was, as first step, aligned to L3 itself. The observables used for the lenses alignment of the BB lenses have been the Newton rings symmetry and their position with respect to the reference position previously defined on CCD#1.

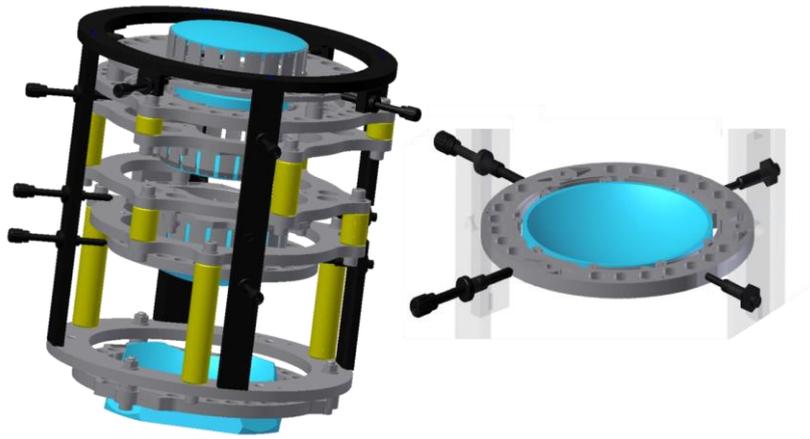

Figure 4: *Left*, in black is shown BB lens centering adjusting system. *Right*, on the right side, in black, the detail of pre-loaded system acting as pistons to keep the lens mount always in contact with the micrometric screws shown on the left side.

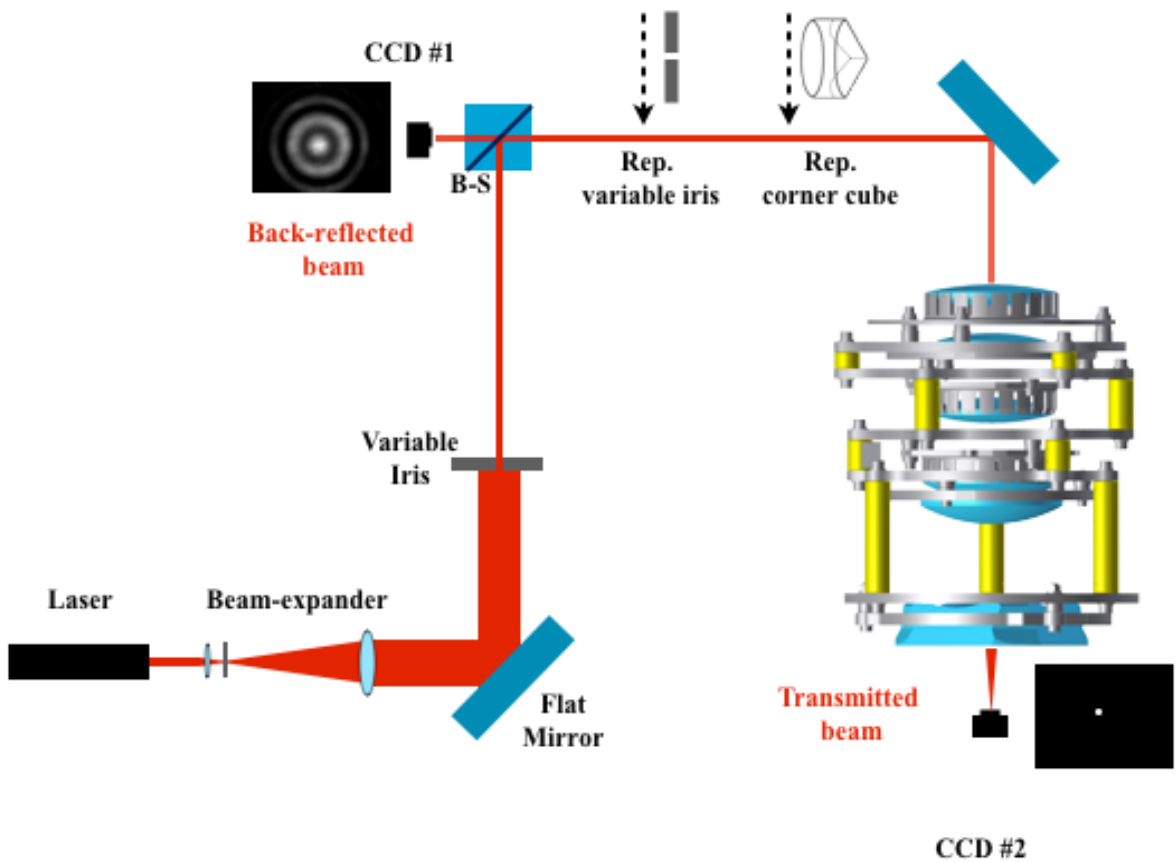

Figure 5: *Left*, setup concept for the BB alignment is shown. A laser beam is expanded using an objective, a spatial filter and a collimating lens. A flat mirror folds the light upward, and a variable iris is used to select the proper diameter, which depends on spots visibility. A beam-splitter sends the light toward a flat mirror which sends the beam toward BB optical axis and allows the light back-reflected from the BB lenses to be observed on a CCD test camera (CCD #1, hereafter). A repositionable corner-cube and variable iris can be inserted in the optical path between the beam-splitter and the flat mirror to define a reference position. Another test CCD (CCD#2, hereafter), located after the BB, collects the transmitted beam, used as reference when a new lens is inserted.

Since they were depending on a combination of centering and tilting of the lens with respect to laser source, these degrees of freedom have been adjusted iteratively. Operatively, the centering of the lens has been adjusted looking at the rings uniformity in the distribution of light, while the tip-tilt has always been corrected centering the back-reflections on the reference. An IDL procedure, where back-reflected profiles were Fourier-filtered, has been used to help the observer measure (with iterations with a user) the position of the rings (see Figure 6). Such a procedure allowed to measure the light intensity of the rings, achieving a standard deviation varying from lens to lens, but always highly inside the tolerances.

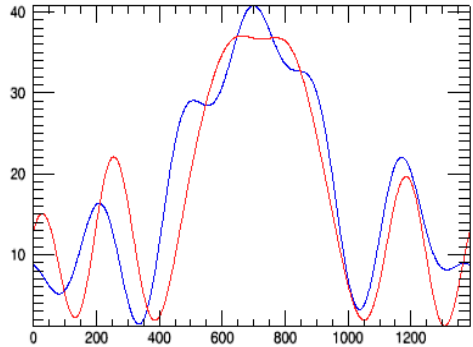

Figure 6: Example of Fourier-filtered back-reflections profiles (L2* is shown). Blue and red profiles correspond to a relative shift of 20 microns of L2*. A shift from red to blue profiles is clearly detectable as an increasing asymmetry of the profile. Units are pixels vs ADU.

After the laser alignment to L3, L2* and L1* where inserted one at a time and aligned looking to the same observables, but centering the lens with the specific tool previously described and shown in Figure 4 and tilted inserting micrometric shims. Before inserting L4, the mechanics of the BB needed to be rotated of 180°. Thus, the first following step was to re-align the laser beam to the already relatively aligned lenses (L1+L2+L3). Then, L4, L5 and L6 were aligned. For L4, though, we could not observe Newton rings, but we used for centering the interference between one surface of L3 with one surface of L4, while L4 quasi-flat surface was considered for the tilt of the lens itself.

During the alignment also transmitted spots positions on CCD#2 were recorded, but finally not used because they were somehow redundant in the devised alignment strategy.

All alignment precision are reported in Table3.

| Lens (insertion order) | Dec. precision | Tilt precision |
|---|---|---|
| L1* (3) | ±10 μm | ±18" |
| L2* (2) | ±20 μm | ±15" |
| L3 (1) | ±32 μm | ±15" |
| L4 (4) | ±15 μm | ±2" |
| L5 (5) | ±10 μm | ±15" |
| L6 (6) | ±15 μm | ±20" |

Table 3: For each lens are reported alignment achieved precisions

A long-term test was performed on the laser beam used as a reference for the alignment, monitoring the spots on the CCD#1 and CCD#2 for 12 hours, a period comparable to the time-scale in which each group of lenses could be aligned. No evident long-time trends emerged and the detected ranges could be traced back to a displacement of the laser beam in the ranges ±5'' and ±15 μm.

In table 4 are reported the alignment indetermination of each of the 6 lenses, computed as a root mean square of the estimated indeterminations. Such indeterminations have to be compared to decenter and tilt tolerances, determined assuming a 10% deterioration of nominal performance of 10%. As reported in Table 4 , the alignment process allowed to reach decenter and tip-tilt precisions inside the tolerances for all lenses, except L3. We remind that laser was aligned to L3, and so, probably, higher accuracies micrometers should have been foreseen for the laser motion.

| Lens | Reference definition effect (tilt) | Laser movement effect (decenter) | Laser movement effect (tilt) | Dec. precision | Tilt precision | Dec. tolerance | Tilt tolerance |
|---|---|---|---|---|---|---|---|
| L1* | ±9.6" | ±15 μm | ±5" | ±18 μm | ±21" | ±30 μm | ±108" |
| L2* | ±10" | ±15 μm | ±5" | ±25 μm | ±18.7" | ±50 μm | ±36" |
| L3 | ±16" | ±15 μm | ±5" | ±35.3 μm | ±22.5" | ±30 μm | ±36" |
| L4 | ±8" | ±15 μm | ±5" | ±21.2 μm | ±9.6" | ±30 μm | ±36" |
| L5 | ±4" | ±15 μm | ±5" | ±18 μm | ±16.3" | ±30 μm | ±36" |
| L6 | ±23.2" | ±15 μm | ±5" | ±21.2 μm | ±31" | ±70 μm | ±216" |

Table 4: For each lens are reported, from left to right, propagation of the reference indetermination on the tilt alignment precision, indetermination due to laser movements (decenter and tilt), final alignment achieved precision (computed as the root mean square of the 3 previous columns and the values reported in Table 3), and, finally decenter and tilt tolerances given for comparison.

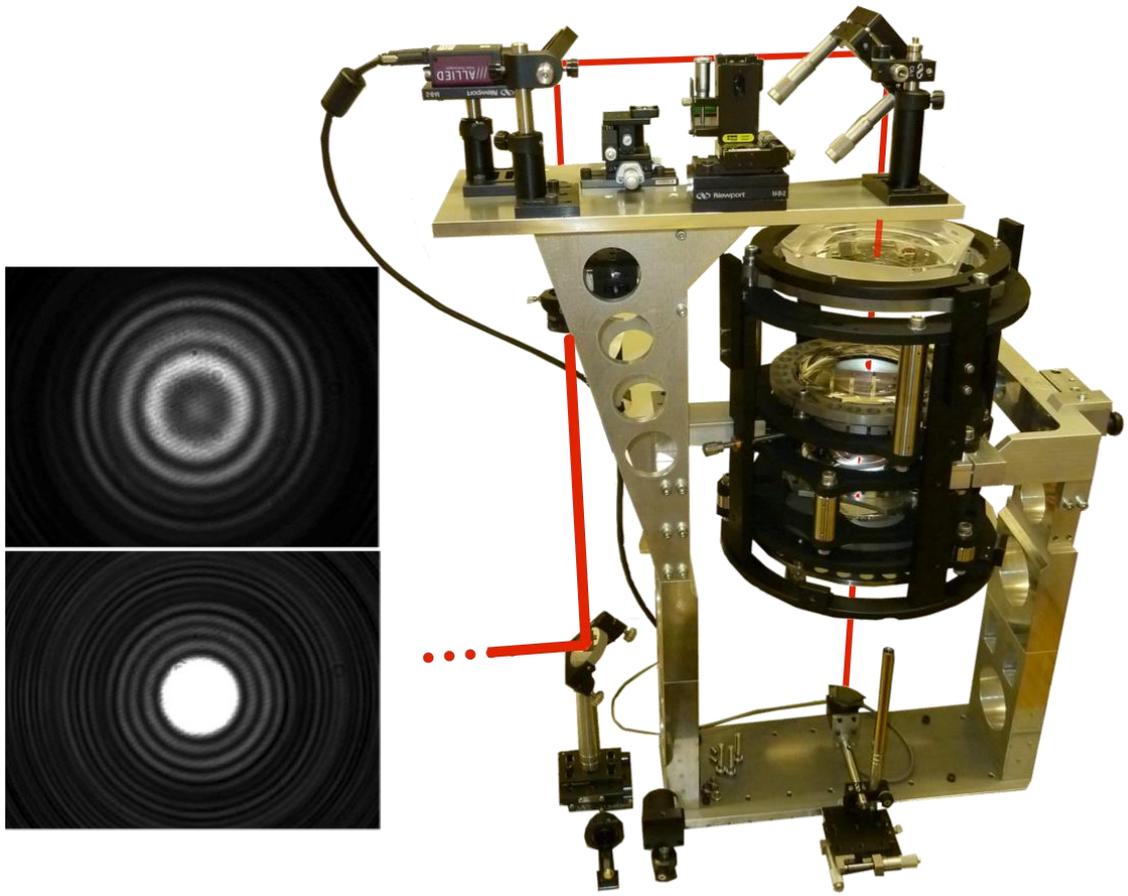

Figure 7: The handling used to hold the BB during the integration. Such structure allows the required 180 degrees rotation of the BB itself, to insert the lenses, starting from L3, always from the top of the structure. On the left are shown the back reflected spots coming from L1*+L2*+L3 (top), and from all lenses -L6+L5+L4+L3+L2*+L1*- (bottom).

# 5. TOU PROTOTYPE RESULTS

After TOU assembly and integration, optical quality test have been performed in Padova in "warm" conditions (20° C, 1 atm) and afterwards at SELEX Galileo premises, in "cold" conditions (-80°C, 0 atm), with the purpose to double-check the on-axis performance in the final working conditions.

BB was dismounted from its vertical handling and positioned on a horizontal handling, designed to be used both in "warm" and "cold" conditions. The handling consisted of 3 V-shaped columns on a common base-plate, where BB rested without any clamping device, in order to avoid introducing any mechanical or optical stress on the BB (see Figure 8).

The problems to be faced due to low testing temperatures and gradient between inner and outer parts of the chamber determined the tests to be performed. Selected tests present the advantage of requiring an input parallel beam, to minimize the entrance windows aberrations due to different internal and external conditions, and to avoid the presence of additional optical components inside the chamber that would have introduced extra-aberrations and required specific mounts.

## 5.1 Interferometric test

A standard interferometric test was performed in "warm" conditions, using the Zygo interferometer, producing a 100mm collimated beam. A F/1.5 transmission sphere was the optical output element to achieve the proper input F/#. In Figure 8 left, is shown the setup: the BB is held on its horizontal handling, between the interferometer and a flat mirror with a certified optical quality better than λ/10 RMS.

The results of the interferometric tests (see Figure 8 right) report values of 1.68 waves PtV and 0.31 waves RMS (lambda = 632.8 nm), values within the expected ones. The alignment of the BB to the interferometer and flat mirror, through back-reflected spots to minimize coma aberration, was extremely hard, and tilting screws of the mirror not precise enough, so we consider the measured wavefront an upper limit.

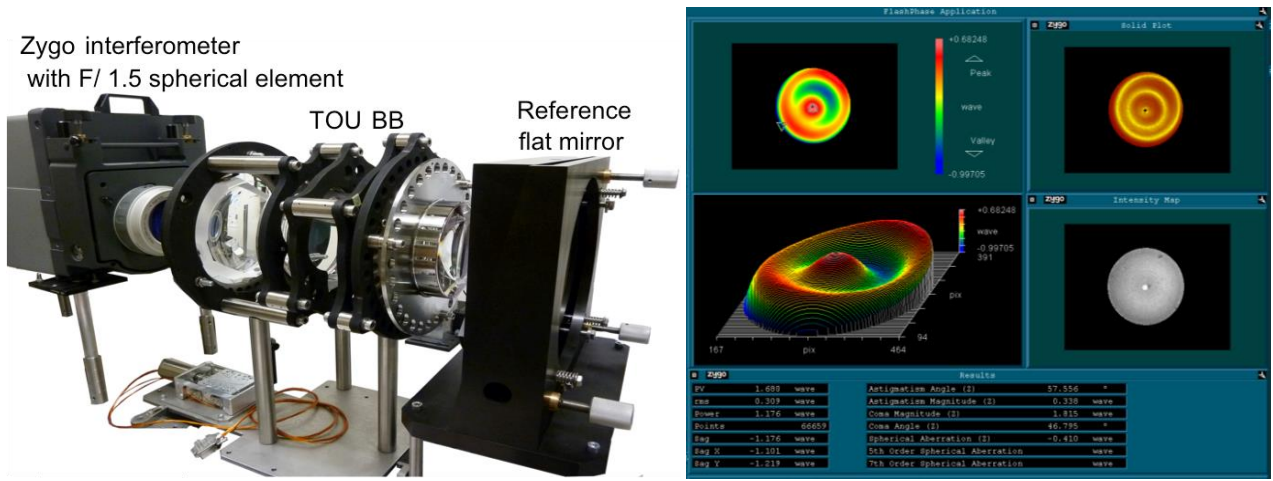

Figure 8: *Left*: interferometric setup (is clearly visible the horizontal handling and the motorized linear stage use to move CCD along optical axis in PSF and Hartmann tests). *Right*: interferometric results: the measured PtV WFE is 1.68 waves (@632,8 nm), while RMS WFE is 0.31 waves.

## 5.2 PSF test

PSF tests were done in monochromatic HeNe light using the interferometer collimated beam and a CCD mounted on a motorized linear stage (visible in Figure 8 left) positioned after the BB. Because of the characteristic of the BB optical design, the PSF shape and dimension had to be measured only on-axis. In "warm" conditions, interferometer beam was 100 mm, while in "cold" it was expanded to 300 mm (BB entrance pupil is 120 mm). In both cases the entrance beam had to be aligned to the BB.

The PSF dimension obtained in cold conditions had a FWHM of 1.4 TOU pixels, a value compatible to have 90% of EE within tolerances, but due to the very high noise the CCD, for more reliable results we point the reader toward the Hartmann test.

### 5.3 Hartmann test

A Hartmann test was performed in order to measure the optical quality of the BB. The results have to be compared with the expected performance, recomputed with final values of lenses, index of refraction, and shimming, shown in Figure 9.

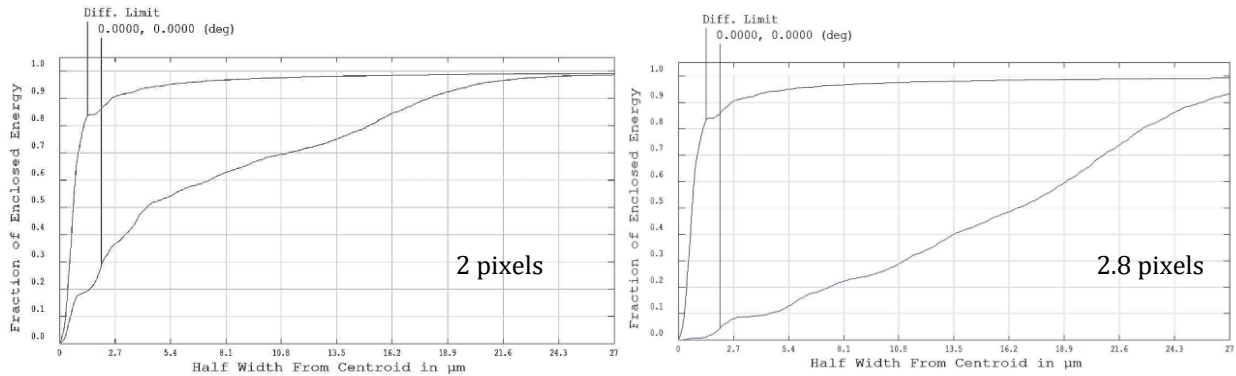

Figure 9: Expected EE for the warm (right) and cold (left) case, 2 and 2.8 TOU pixels respectively.

A Hartmann mask has been designed to sample with 76 apertures a 120 mm beam (BB entrance pupil size). To perform image analysis three sets of 100 images were taken at three CCD locations along the optical axis, each one separated by 1 mm, reached moving the camera with the linear stage. For each set, the average of the 100 images has been computed, to minimize the contribution of bad pixels and electronic noise, obtaining three processed frames (two of them are shown in Figure 10 left). Afterward, centroids of the 76 spots, for each frame, have been computed as the center of mass of the light distribution of each spot, enclosed in a box having an arbitrary size. We point out that no further optimizations of the centroids computations (background subtraction, box size, outliers discard, systematic effects analysis…).

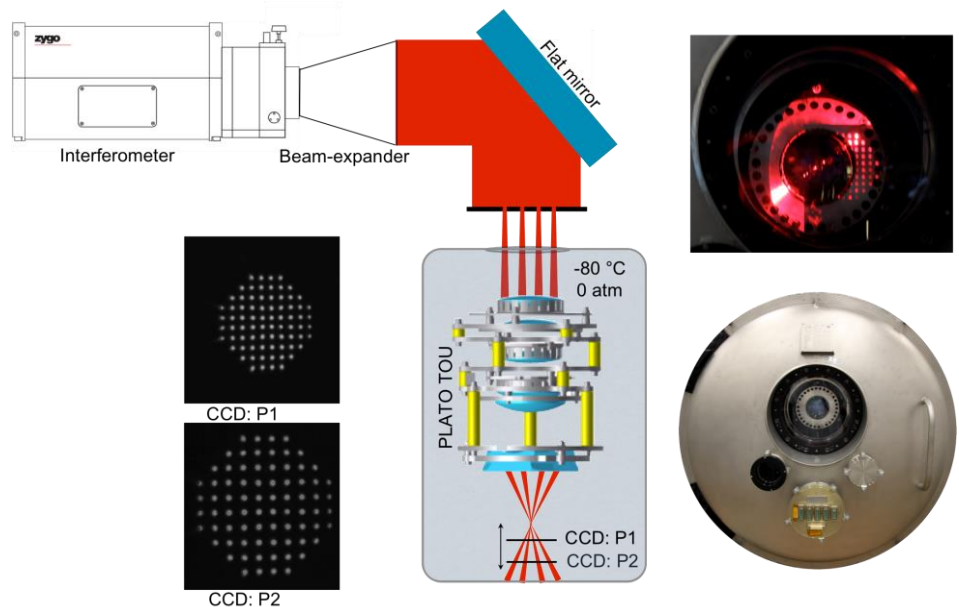

Figure 10: The setup for the Hartmann test is shown. A collimated beam coming from the interferometer, expanded to 300 mm, was folded by a flat mirror toward the cryo-vacuum chamber, where BB had been positioned (see bottom right picture). In the collimated beam a Hartmann mask was positioned (its shadow is visible on the top right picture). The CCD had to be moved along the optical axis to retrieve the focus position and the EE. On the left are shown the two images of the spots produced by the Hartmann mask at a relative distance of 2mm in the optical axis direction.

The centroids of the two frames taken at a distance of 2mm have been used to compute the parameters of the lines representative of the 76 beams. Such parameters allow reconstructing the "rays" distribution at any focal distance, as shown in Figure 11 (*left*).

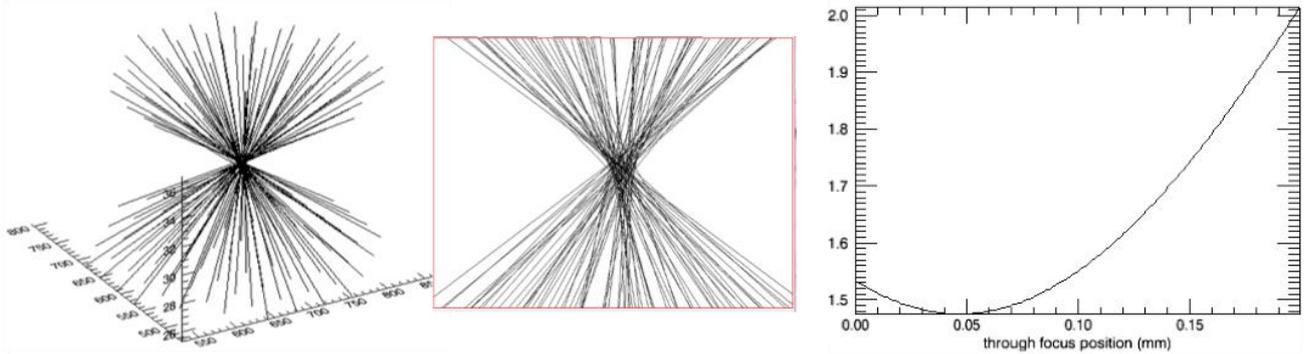

Figure 11: The 3D reconstruction of the rays defined by the couples of spots centroids computed on the two images. *Left*: the effective 3D reconstruction, x and y values are TDS pixels (pixel size=7.45μm), while z values are the position through focus, in mm. *Center*: zoom of the 2mm range, centered on the focal plane position, the expected spherical aberration, due to the use of L1* and L2*, is clearly visible. *Right*: through focus RMS radius of the extrapolated centroids.

The indetermination, intrinsic in the centroid-computing procedure, has been quantified comparing the extrapolated positions of the rays in the middle of the range between the two images with the centroids measured on the third frame. The focal plane position has been computed as the through focus position in which the RMS radius of the 76 centroids was minimized (see Figure 12). The enclosed energy has been measured on the extrapolated positions of the 76 rays, in correspondence of the focal plane. Cold focal plane position differs from the warm one of about 1.4 mm, instead of the 1.95 mm expected (see Section 2.2), but this difference is within expected values, because CCD was on a different mount than BB and the two could move differentially in the "warm" to "cold" transition.

Figure 13 summarizes the results in terms of Enclosed Energy (EE):

- in the warm case, the 90% EE corresponds to a diameter of 5.8 TOU pixels. The estimated intrinsic error is 0.77 TOU pixels
- in the cold case, the 90% EE corresponds to a diameter of 4.1 TOU pixels. The estimated intrinsic error is 0.47 TOU pixels

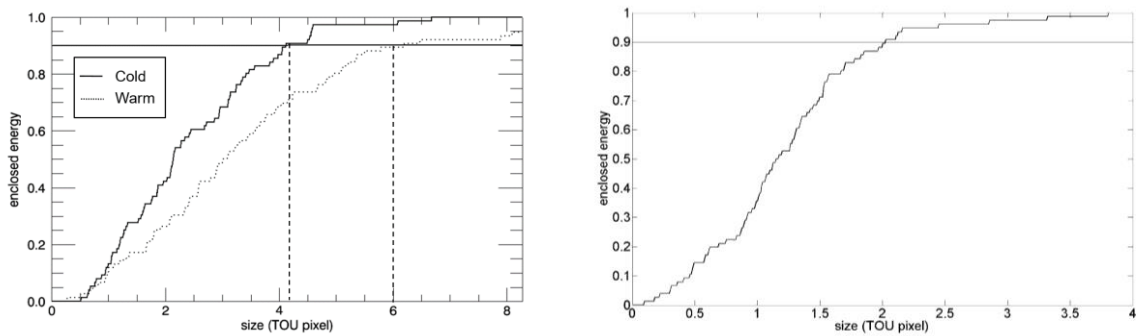

Figure 12: Estimated enclosed energies in the best focal positions, in warm (dotted line) and cold (solid line) conditions. Solid straight line represent the 90% EE level. Right: Estimated enclosed energy in the best focal position, obtained with a finer data reduction. Solid straight line represent the 90% EE level.

Even though the values are higher than expected, the retrieved enclosed energies show that the system maintains the alignment in the transition from warm to cold conditions and, moreover, that the performance improves in the transition from the ambient environment (where the alignment has been performed) to the cryo-vacuum one of an amount comparable to the expected ones.

The already defined intrinsic error, computed on the intermediate image, allows an independent and safe estimation of the centroids indetermination, having the best approach established on scientific basis. Careful centering of the Hartmann spots and a better definition of the "best" focal plane are two examples of areas where data analysis could be improved to be less sensitive to a relatively poor quality of the Hartmann spots and to assess a better understanding of the breadboard behavior.

However, this (and further) kind of analysis is beyond the limit of this paper, which was solely intended to show the feasibility of the proposed alignment approach. Nevertheless, Figure 14 shows an example of data reduction (in which all the data collected are treated) where the retrieved optical quality (90%EE in 2.03 TOU pixels) is consistent with the targeted one.

## 6. $CaF_2$ PERFORMANCE

As previously described, $CaF_2$ has been chosen in the final design because of the great advantages presented for the system. However, information over the response of this material to the use in space environment is limited. Only in recent years, real tests have been performed on $CaF_2$ [6], where $CaF_2$ plane parallel polished blanks, (20 mm diameter, 5 mm thickness) were exposed to high variations of temperature and UV and gamma radiations, concluding that optical and mechanical performances were not affected.

Key in this paper is the analysis of $CaF_2$ L3 lens performance during prototype tests, to assess whether this brittle material suffers in TOU space working conditions (0 atm, -80°C) and results coming from tests performed over two $CaF_2$ blanks, where we wanted to double-check if any degradation due to expansion and contraction of the material would appear after thermal cycle, and to study $CaF_2$ response to vibrations (simulating launch stress).

$CaF_2$ lens installed in the breadboard survived without any visible damage or deterioration during alignment at ambient temperature and during thermo-vacuum tests where temperatures cycled from 20 °C to -80°C. Moreover, optical test confirmed the simulated performance going from warm to cold condition.

Two $CaF_2$ blanks were ordered from Korth-Kristalle GmbH, in order to perform specific tests. The samples are non-polished parallel-plane disks with 26 mm of thickness and 120 mm diameter, similar to the flight model lens dimensions (23 mm thickness, 116 mm diameter). They are fixed on a barrel different from the final one, see Figure 15. The lens is glued on the barrel with the same glue and process as planned for flight model, but connection points have been chosen to maximize the thermal contact between the barrel and the facilities. In principle, due to rough surfaces and the presence of edges, these blanks are even more fragile than final lenses, because edges are likely to become starting points of cracks.

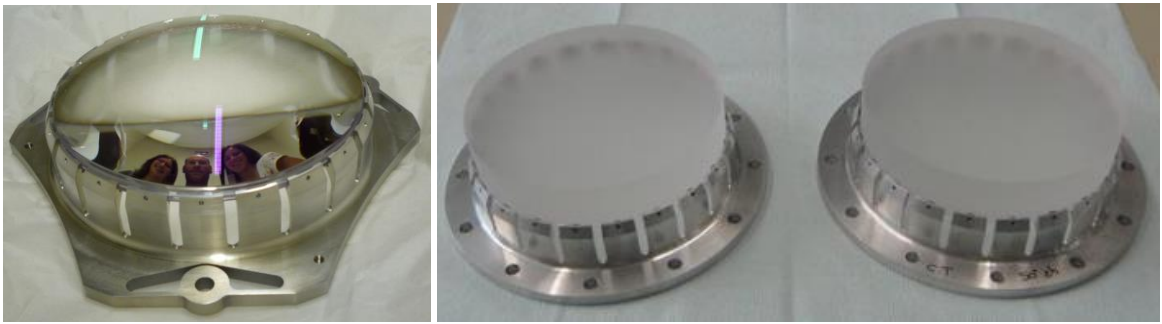

Figure 13: Comparison of lens mounts for final $CaF_2$ (left) lens and lens mounts for $CaF_2$ blanks (right)

Both blanks have been glued to their mounts by SELEX Galileo; one has been cured, with an oven at 50°C and the other one, instead, was left at room temperature. No problems arose during this operation.

The first blank was then sent to University of Bern, where vibration tests were performed. During tests, two tri-axial accelerometers were used to perform tests on x,y,z axes at various resonance frequencies, one was positioned on the lens and the other one used for a closed-loop monitoring (see Figure 16). Visual inspection after tests confirmed that no damage deterioration or evident changes in mechanical properties were observed. Anyway during this test, there was a confirmation on the extreme care needed to work with this lens material. In fact, while dismounting the lens, an edge was slightly touched with a head-ball screwdriver and immediately two cracks of a few mm for a few mm of deepness propagated through the material. The cracked area was removed for the following tests, performed at CNES Toulouse facility, where the blank was tested to verify the strength of material in PLATO temperature range.

The lens was tested in thermo-vacuum, with temperatures cycling from 45 °C to -100°C, and a gradient of 0.2 °C/min. Visual changes confirmed no deterioration after the test, even in areas, which were previously identified as weak because of small defects in the bulk of the lens.

Additionally, we have to report that not foreseen extra-vibration and thermal cycling tests, with a very fast °C gradient, were performed due to very high external temperature in June during the transportation (by car) from Firenze to Padova.

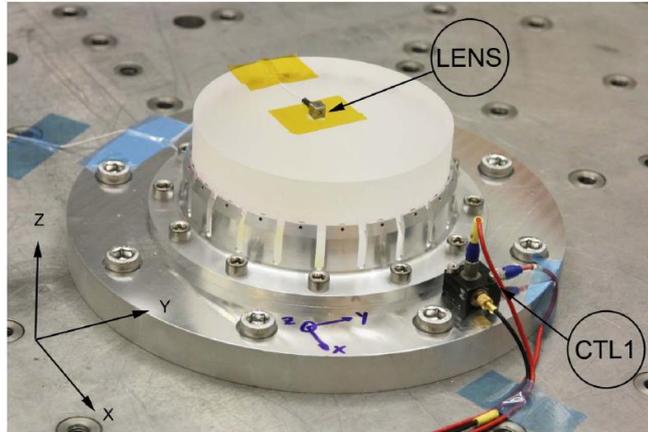

Figure 14: Setup used for the vibration test. Indicated are the two tri-axial accelerometers, one (LENS) used to verify lens reaction to vibrations and the other (CTL1) for closed-loop control.

## 7. CONCLUSIONS

In conclusion, alignment of Plato TOU prototype at Padova Laboratories, confirmed the feasibility of the alignment of the TOU in a time-slot estimated in 2-3 days, a reasonable amount of time if this had to be performed for the 32 TOUs by industries.

Both optical and mechanical components behaved extremely well, allowing to meet alignment requirements in warm, maintaining the alignment also in the transition from warm to cold, where the performance improved of an amount comparable to the expected one. No problem arose with L3, $CaF_2$ close-to-pupil lens, confirming the possible choice of this material. The only small problem was encountered for L6 coupling with its mount, which caused some stress on the attachment points. We remind it was the only mount with only 4 connection points, had a squared shape and was bigger in size than all the other lenses.

Moreover, extremely important also for future space projects, we have intensely tested $CaF_2$ blanks, which have revealed no evident damage nor sign of deterioration even after intensive tests for vibrations and thermal variations, simulating working conditions (0 atm, -80°C) and launch stress.

The idea behind all these tests was to verify the behavior of the blanks, more fragile than a polished lens, due to the presence of edges, possible cracks propagators, to be confident, in case of success, to affirm that $CaF_2$ lens installed on a specific flexible quasi-isostatic mount with the proper material, if treated with an absolute care, can be considered a safe material.

CaF2 lenses and blanks have demonstrated to be resistant to mechanical and thermal shocks and we therefore affirm that their use could be considered in future optical design for space systems.

# REFERENCES

[1] Czichy R. H,. "Optical Design and technologies for space instrumentation", Proc. SPIE 2210, 420-433 (1994)
[2] Schott Lithotec Catalogue
[3] Henson T.D. and Torrington G. "Space radiation testing of radiation of resistant glasses and crystals," Proc. SPIE 4452, 54-65 (2001)
[4] Al-Jumaily G. A. "Effects of radiation on the optical properties of glass materials," Proc. SPIE 1761, 26-34 (1992)
[5] Yan J.,Syoji K and Tamaki J., "Crystallogaphic effects in micro/nanomachining of single-crystal calcium fluoride", J. Vac. Sci. Technol, B 22 (2004)
[6] Fernandez-Rodriguez F, Alvarado C. G., Nunez A., Alvarez-Herrero A., "Analysis of optical properties behaviour of clearceram, fused silica and $CaF_2$ glasses exposed to simulated space conditions", Proc. ICSO (2010)
[7] Catala, C, PLATO Consortium, "PLATO: PLAnetary Transits and Oscillations of stars", Journal of Physics: Conference Series, Volume 118, Issue 1, id. 012040 (2008)
[8] Plato Definition Study Report (Red Book), ESA (2011)
[9] Magrin D., Munari M., Piazza D., Ragazzoni R., Arcidiacono C., Basso S., Dima M., Farinato J., Gambicorti L., Gentile G., Ghigo M., Pace E., Pagano I., Piotto G., Scuderi S., Viotto V., Zima W., Catala C., "PLATO: detailed design of the telescope optical units", SPIE Proc. 7731, 773124-5 (2010)
[10] PLATO Team, PLATO-INAF-TOU-REP-0012 Issue 0.1 Rev.3 (2011)